\documentclass[sigconf, authorversion, screen]{acmart}  

\usepackage[utf8]{inputenc}        
\usepackage[nameinlink]{cleveref}  

\copyrightyear{2023}
\acmYear{2023}
\setcopyright{rightsretained}
\acmConference[ITiCSE 2023]{Proceedings of the 2023 Conference on Innovation and Technology in Computer Science Education V. 1}{July 8--12, 2023}{Turku, Finland}
\acmBooktitle{Proceedings of the 2023 Conference on Innovation and Technology in Computer Science Education V. 1 (ITiCSE 2023), July 8--12, 2023, Turku, Finland}
\acmDOI{10.1145/3587102.3588862}
\acmISBN{979-8-4007-0138-2/23/07}

\makeatletter
\gdef\@copyrightpermission{
   \begin{minipage}{0.3\columnwidth}
     \href{https://creativecommons.org/licenses/by-sa/4.0/}{\includegraphics[width=0.90\textwidth]{figures/4ACM-CC-by-sa-88x31.eps}}
   \end{minipage}\hfill
   \begin{minipage}{0.7\columnwidth}
     \href{https://creativecommons.org/licenses/by-sa/4.0/}{This work is licensed under a Creative Commons Attribution-ShareAlike International 4.0 License.}
   \end{minipage}
   \vspace{5pt}
}
\makeatother

\renewenvironment{quote}{\list{}{\leftmargin=1.3em\rightmargin=1.3em}\item\relax\hspace{-0.2em}\ignorespaces}{\unskip\unskip\endlist}

\begin{document}


\title[Want to Raise Cybersecurity Awareness? Start with Future IT Professionals.]{\texorpdfstring{Want to Raise Cybersecurity Awareness? \\ Start with Future IT Professionals.}{Want to Raise Cybersecurity Awareness? Start with Future IT Professionals.}}

\author[L. Kraus]{Lydia Kraus}
\orcid{0000-0002-1387-3578}
\affiliation{%
   \institution{Masaryk University}
   \city{Brno}
   \country{Czech Republic}
}
\email{lydia.kraus@mail.muni.cz}

\author[V. Švábenský]{Valdemar Švábenský}
\orcid{0000-0001-8546-280X}
\affiliation{%
   \institution{Masaryk University}
   \city{Brno}
   \country{Czech Republic}}
\email{valdemar@mail.muni.cz}

\author[M. Horák]{Martin Horák}
\orcid{0000-0002-1835-6465}
\affiliation{%
   \institution{Masaryk University}
   \city{Brno}
   \country{Czech Republic}
}
\email{horak.martin@mail.muni.cz}

\author[V. Matyáš]{Vashek Matyáš}
\orcid{0000-0001-7957-7694}
\affiliation{%
   \institution{Masaryk University}
   \city{Brno}
   \country{Czech Republic}}
\email{matyas@fi.muni.cz}

\author[J. Vykopal]{Jan Vykopal}
\orcid{0000-0002-3425-0951}
\affiliation{%
   \institution{Masaryk University}
   \city{Brno}
   \country{Czech Republic}
}
\email{vykopal@ics.muni.cz}

\author[P. Čeleda]{Pavel Čeleda}
\orcid{0000-0002-3338-2856}
\affiliation{%
   \institution{Masaryk University}
   \city{Brno}
   \country{Czech Republic}
}
\email{celeda@ics.muni.cz}

\begin{abstract}
As cyber threats endanger everyone, from regular users to computing professionals, spreading cybersecurity awareness becomes increasingly critical. Therefore, our university designed an innovative cybersecurity awareness course that is freely available online for students, employees, and the general public. The course offers simple, actionable steps that anyone can use to implement defensive countermeasures. Compared to other resources, the course not only suggests learners what to do, but explains why and how to do it. To measure the course impact, we administered it to 138 computer science undergraduates within a compulsory information security and cryptography course. They completed the course as a part of their homework and filled out a questionnaire after each lesson. Analysis of the questionnaire responses revealed that the students valued the course highly. They reported new learning, perspective changes, and transfer to practice. Moreover, they suggested suitable improvements to the course. Based on the results, we have distilled specific insights to help security educators design similar courses. Lessons learned from this study are relevant for cybersecurity instructors, course designers, and educational managers.
\end{abstract}

\ccsdesc[500]{Applied computing~Education}
\ccsdesc[100]{Security and privacy}

\keywords{cybersecurity education, course evaluation, information security awareness, computer science undergraduates}

\maketitle
\balance            

\section{Introduction}
Protecting oneself from security and privacy threats in cyberspace is challenging. 
IT-knowledgeable users are thereby an important information source for users without IT background~\cite{redmiles2016learned}. 
However, where do these IT-knowledgeable users learn about security advice and behaviors?
The literature indicates that there is no unified source for security advice online; the advice seems to be spread across the Internet, and opinions about which advice should be prioritized diverge among lay users and experts~\cite{redmiles2020comprehensive}. 
Our university offers a unified source of advice: the Cybercompass, which is a freely available online resource for students, employees, and the wider public~\cite{muni2019cybercompass}. The course consists of five lessons: \textit{Security of devices}, \textit{Passwords}, \textit{(Cybersecurity) Self-defense}, \textit{Secure communication}, and \textit{Incident reporting}. 

To raise cybersecurity awareness among IT-knowledgeable users, we included the Cybercompass in a compulsory introductory course to information security and cryptography (ISC) for computer science students and evaluated their experiences. 
Thereby, we assigned them a homework to explore the course. 
After each of the five online lessons, students filled in a questionnaire examining their overall impression of the lesson, its usefulness, comprehensibility, and difficulty. 
We further asked whether they learned something new, whether taking the Cybercompass changed their view on everyday cybersecurity, and whether they would recommend the Cybercompass to others, such as family, non-university friends, fellow students, or colleagues.
Our work yields two key contributions:

\begin{enumerate}
     \item \textit{We evaluate the effects of including the Cybercompass material into introductory security courses.} Our results show that students valued the Cybercompass highly. They reported new learning, changes in their perspective, and transfer to practice. 
     \item \textit{We release the Cybercompass.} The course is freely available online~\cite{muni2019cybercompass} and can thus serve as an inspiration for educators who plan to design a similar course.
\end{enumerate}

As a result of our positive experiences, we encourage other teachers to consider including practical cybersecurity hints and defensive countermeasures as covered in the Cybercompass into introductory security courses. This will improve awareness and cybersecurity best practices in the higher education environment and beyond. 

\section{Related Work}

\subsection{Cybersecurity Threats in the Higher Education Sector}
Higher education institutions have become attractive targets during the last years, as several data breaches and surveys indicate~\cite{rezgui2008information,przyborski2019cyberworld}.
As of 2017, the number of data breaches at higher education institutions doubled, and \texttt{.edu} email addresses continue to constitute a popular target for hackers~\cite{przyborski2019cyberworld,bolkan2017education,digital2017cyber}. 
As of 2018/2019, 72\% of higher education institutions consider phishing and social engineering the top threat they are facing~\cite{chapman2020cyber}. 
Ransomware/malware and unpatched security vulnerabilities rank second and third~\cite{chapman2020cyber}.
While in the past, only a third of higher education institutions offered security training for students and staff~\cite{kvavik2003information}, the numbers increased up to 80\%~\cite{chapman2020cyber}. 
Bongiovanni~\cite{bongiovanni2019least} reviewed the literature on information security management in higher education and concluded that the topic is ``highly under-investigated''. 

\subsection{Students' Information Security Awareness}
While often online, students were shown to lack information security awareness, particularly when they enter higher education~\cite{korovessis2013information,jisc2020information}.
Data show a rise of scamming emails targeting students at the beginning of every academic year~\cite{jisc2020information}. 

North et al. surveyed 465 students in introductory computer technology courses at different US universities. Most participants demonstrated a satisfactory awareness of computer security and ethics~\cite{north2006computer}.
Yet, between 20\% and 52\% of participants had knowledge gaps in specific areas of computer security.

Muniandy et al. assessed cybersecurity behavior of 128 students in the categories of malware, password usage, phishing, social engineering, and online scamming~\cite{muniandy2017cyber}. They found that the reported behavior was unsatisfying in several categories. About 30\% of students were unsure about the status of their antivirus software, and almost 30\% indicated that they would be willing to download material from insecure websites. Similarly, about 50\% of the students did not follow safe password practices.

Sheng et al. conducted a survey on the susceptibility to phishing with 1001 participants. They found that younger participants between the ages of 18 and 25 were more susceptible to phishing than other age groups~\cite{sheng2010falls}. 
Lastly, Matyas et al. investigated students' security behavior over the course of several years and found that secure behavior improved, despite less and less students reading the university's security directive \cite{matyas2021even}. 

\subsection{Where Do Students Receive Essential Cybersecurity Training?}
Kim conducted a survey with 68 undergraduate and graduate students about exposure to information security awareness training~\cite{kim2014recommendations}.
Although many students in the study understood the need for information security awareness training, most respondents did not participate in training at the university or work~\cite{kim2014recommendations}. 
Similarly, a CDW-G survey showed a discrepancy between the training that students ought to receive and the training that students actually receive~\cite{lestch2017college}: 82\% of IT professionals said that students need to engage in cybersecurity training at least once a year, yet, only 35\% of students said that was required of them. 

In general, students learn about cybersecurity from various sources such as websites and media rather than from dedicated training~\cite{kim2014recommendations}.
Yet, learning by security advice from the open web has its pitfalls.
Redmiles et al. evaluated advice sources and their quality on the web and identified 374 pieces of unique advice~\cite{redmiles2020comprehensive}. 
They found that a vast majority (89\%) of advice is considered useful by users and experts alike~\cite{redmiles2020comprehensive}. 
However, the common problem is that users and experts struggle to prioritize these advice pieces~\cite{redmiles2020comprehensive}. 
Moreover, many representative security awareness websites do not offer a structured way of conveying advice to end users~\cite{korovessis2017toolkit}. 

\subsection{Security Awareness Training Outside Academia}
The interest in cybersecurity training is high even beyond the academic environment. 
Ricci et al. surveyed more than 200 adults and found that most of them would be interested in a cybersecurity seminar, especially if the employer would pay for it~\cite{ricci2019survey}. 
Yet, the willingness to spend time and money for such a seminar is limited: the desired seminar length was rated 1 to 1.5 hours, and the desired costs were rated \$20 on average, with 40\% of participants not willing to pay at all~\cite{ricci2019survey}.
Regarding the format of cybersecurity education seminars, Ricci et al. further found that more than two-thirds of surveyed participants prefer some form of online education as part of a seminar~\cite{ricci2019survey}.
Given the facts described above, it is important that organizations and institutions offer free and efficient online cybersecurity education for everyone.  

\section{The Evaluated Cybersecurity Awareness Course}

Our university offers a structured cybersecurity awareness course. The Cybercompass is an extracurricular activity in the form of educational material presented on a website. It is open-source and freely available for anyone on the Internet~\cite{muni2019cybercompass}. 
The most valuable features of the course are two-fold:
First, it offers and prioritizes security advice.
Second, the advice presented on the website is complemented with reasoning: it explains both, the \textit{what} of proper security behavior and the \textit{why}.

\subsection{Course Design}
The Cybercompass was designed by a multidisciplinary team in 2019. The main challenge was to prepare an easily accessible course for all users, which will positively influence their security behavior. The team used the Design Thinking methodology~\cite{ideo2021design}, working with users from different target audiences during the design process. Specifically, the team focused on simplicity of language, chunking of topics into lessons, appealing visual style, and intuitive interactions with the website. Also, the content of the course was reduced to a reasonable minimum (1--2 hours in total) with the goal to provide information that can help users practically perform basic cybersecurity measures and influence their behaviors. 

\begin{figure*}[!ht]
    \includegraphics[width=0.815\textwidth]{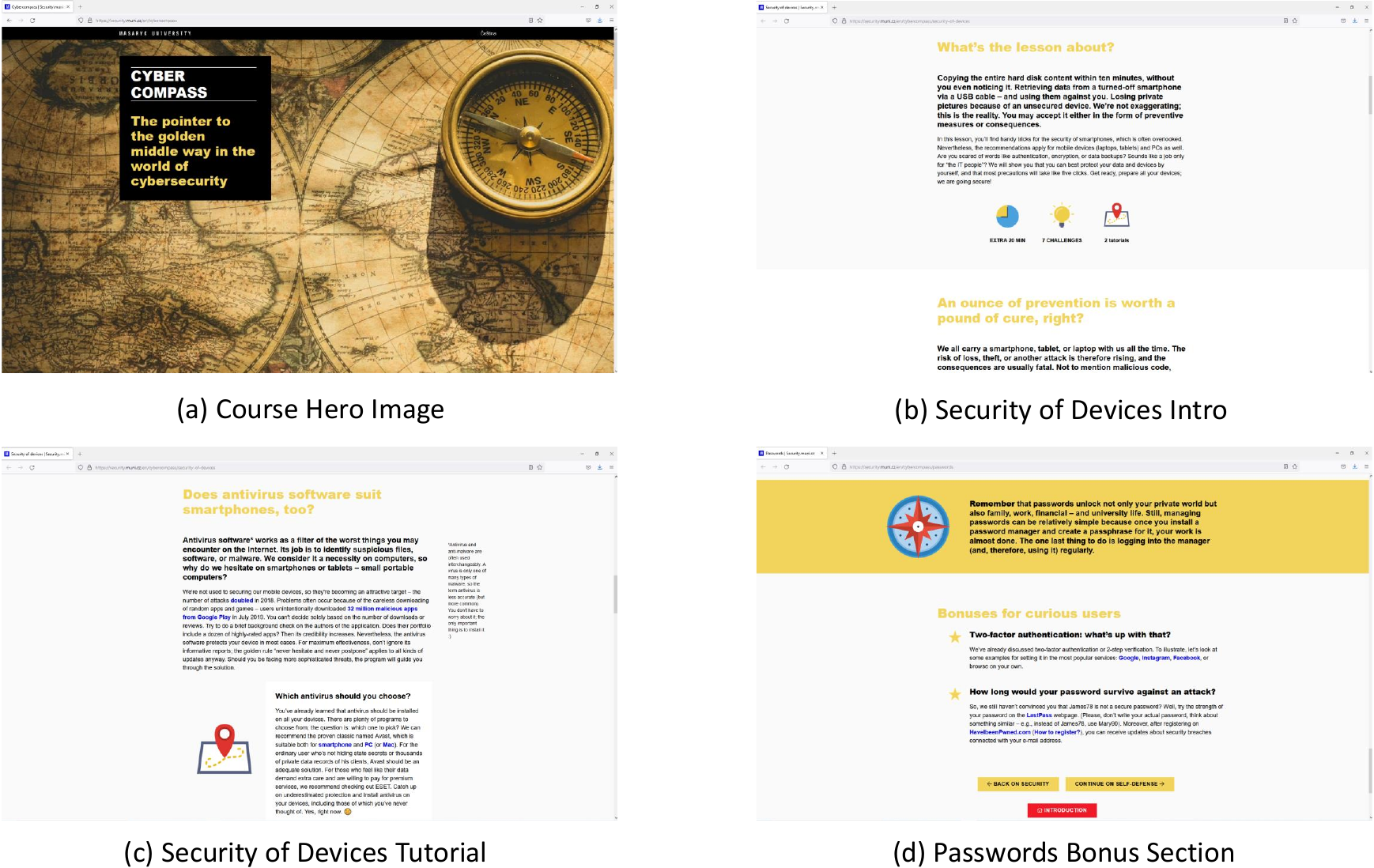}
    \caption{The Cybercompass and its structure (example screenshots) \cite{muni2019cybercompass}.}
    \label{fig:csac-screenshots}
\end{figure*}

Specific security measures were identified from different sources. First, the team studied numerous sources that deal with information security awareness (such as~\cite{ncsa2021how,cyberdegrees2021internet,grok2021schools,berkeley2021information,cleveland2021technology}). Second, the team discussed security measures with members of two Computer Security Incident Response Teams (CSIRTs). Finally, the team also included three members who focused on information security awareness. 
All identified security measures were then internally evaluated in terms of their suitability for course users. Afterward, the final selection was checked with the members of CSIRTs and tested with users. The prototype of the course was iteratively improved based on insights from users.

\subsection{Course Structure and Content}
The course consists of five lessons. 
Each lesson takes between 15 and 30 minutes to complete.
The lessons (except for \textit{Incident reporting}) contain text with factual information about cybersecurity threats and tutorials on how to better protect oneself in cyberspace, together with examples of and links to protective tools (such as a password manager and anti-malware software). 
At the end of each lesson, there is a bonus section for curious users. 
Example screenshots of the Cybercompass are provided in Figure~\ref{fig:csac-screenshots}.

The course offers a variety of topics, covered in five lessons as described above: \textit{Security of devices}, \textit{Passwords}, \textit{(Cybersecurity) Self-defense}, \textit{Secure communication}, and \textit{Incident reporting}.

\begin{enumerate}
    \item \textbf{Security of devices:}  explains the importance of antivirus software on smartphones and PCs, encompasses a step-by-step tutorial for online and offline backups, and encourages the use of screen locks, device encryption, anonymous browsing mode, and others. Furthermore, it advices the rapid installation of updates. 
    \item \textbf{Passwords:} teaches secure password creation with pass\-phrases and encourages the use of a password manager. This is accompanied by a tutorial about how to install and set up a particular password manager. It discourages bad password practices and closes with a bonus section about two-factor authentication (2FA) and password strength-checking.
    \item \textbf{Cybersecurity self-defense:} provides a phishing guide, a phishing quiz, and information on how to avoid dangerous websites. It raises awareness about users' visibility in public, password-protected, and virtual private networks (VPNs). 
    \item \textbf{Secure communication:} introduces learners to the benefits of Eduroam (an international roaming service for higher education institutions), including a tutorial on how to install the related configuration software. It further teaches learners to use the university VPN and information system features for secure file sharing. It links to guides for obtaining personal certificates for e-mail encryption and signing. It closes with a bonus section that links to the university IT services website and further useful applications.
    \item \textbf{Incident reporting:} provides the learners with a step-by-step guide of how to report a cybersecurity incident, accompanied by relevant contact information and a picture of members of the university CSIRT.
 \end{enumerate}

\section{Evaluation Methodology}

To investigate how students perceive the course in different dimensions, we administered an online questionnaire, asking students to evaluate the course, posing questions related to the course outreach, and exploring its impact.  
We designed the study in an open manner with no preset hypotheses to capture the unique and disparate issues arising from interacting with the course. 
The evaluation presented was done by an independent team not involved in the course design. Yet, this paper is authored by members of both teams -- the evaluation and the course design team.
The evaluation ran in spring 2021.

\subsection{Setting}
To find out more about students' prior exposure to security advice and to confront them with a unified resource of advice, we included the Cybercompass into a homework assignment of a compulsory introductory course to information security and cryptography (ISC). This course is taught yearly at our Faculty of Informatics. 
It is mandatory for undergraduate students of computer science in their second year and encompasses between 250 and 300 participants every year.
The ISC course consists of 12--13 lectures with accompanying seminars and 5--6 homework assignments over the course of the semester.

Students receive up to six points for finishing homework assignments. 
Each point contributes 1\% to the total grade from the course. 
To finish the ISC course successfully, students need to achieve at least 50\%.
We decided to reward them with 1.5 points (1.5\%) for taking the Cybercompass and answering related questionnaires and 3.5 points (3.5\%) for creating new educational material that could potentially be used to enhance the course in the future.

\subsection{Study Design}
We asked students to proceed through the course lesson by lesson. 
After each Cybercompass lesson, we had students fill in a questionnaire that encompassed these issues: 

\begin{itemize}
    \item \textbf{Course evaluation:} What is students' overall impression of each lesson? Do they find the lessons useful, comprehensible, or difficult? Do students learn something new in each lesson? Do students use the suggested activities and tools? Do students read the bonus material at the end of each lesson?
    \item \textbf{Course outreach:} Have students heard about the course before? If so, where did they learn about it?
    \item \textbf{Course impact:} Does the course change students' view on preparedness and education regarding everyday cybersecurity (cf.~\cite{breitinger2021first})? If so, how? Would students recommend the course to people in their circles, such as family, friends, fellow students, and colleagues?
\end{itemize}

Each questionnaire further contained three questions about the lesson's content to check whether students had worked with the material. 
To harvest further insights into students' reasoning, the questionnaire featured both, closed and open-ended questions.
Students could opt-out of their data being used for research purposes; of 219 students who were enrolled in the course, 138 participated in the study.

Pilot testing before deployment did not indicate any major issues. 
The study did not request ethics approval because no personal data of students were collected. The questionnaire is available as a supplementary material to the paper at \cite{kraus-cybercompass-questionnaire}.

\subsection{Data Analysis}
We conducted statistical analyses to investigate the answers to the closed-ended questions. To do so, we transformed the answer scales into numerical values. Details are described in Section~\ref{sec:results}. Answers to open-ended questions were analyzed using qualitative data analysis techniques. We first performed open coding and identified themes by question and by lesson. Thereafter, we looked for re-occurring patterns across lessons, i.e., across the whole course (axial coding).

\subsection{Computer Science Student Population}
In spring 2021, 219 students were enrolled in the course. 176 were male and 43 were female. As most students enter our faculty directly after graduation from high school, we estimate their average age to 20--22 years. 154 students were enrolled into the computer science bachelor program, 52 into the software development program, and 13 into other programs. 
University-internal survey reveals that 83.3\% of graduates from our faculty work after graduation in the field of information and communication activities, with, for example, programming, consulting, management of computer equipment, and other activities \cite{nekuda2021}.

\section{Results}
\label{sec:results}
We now interpret the answers of 138 students who participated in the evaluation of the Cybercompass.
Only few of them (1--5 for each lesson) answered two or more reading check questions incorrectly; their answers were thus excluded from the respective questionnaires.

\subsection{Successes}
\label{sec:insights-successes}

\subsubsection*{Cybercompass lessons are perceived as positive and useful.}
We had students evaluate their overall impression of each lesson on a Likert scale from 1 (very negative) to 5 (very positive). Each lesson was perceived on average as \textit{somewhat positive} with a tendency to \textit{very positive} (\textit{Security of devices}: $M=4.22$, $SD=.76$; \textit{Passwords}: $M=4.26$, $SD=.83$; \textit{Self-defense}: $M=4.36$, $SD=.74$, \textit{Secure communication}: $M=4.17$, $SD=.82$, \textit{Incident reporting}: $M=4.33$, $SD=.81$). Similarly, the usefulness of each lesson was rated on average as \textit{somewhat useful} with a tendency to \textit{very useful} (\textit{Security of devices}: $M=3.22$, $SD=.71$; \textit{Passwords}: $M=3.34$, $SD=.75$; \textit{Self-defense}: $M=3.34$, $SD=.80$, \textit{Secure communication}: $M=3.37$, $SD=.72$, \textit{Incident reporting}: $M=3.32$, $SD=.86$). Note that the usefulness scale encompassed four items from 1 (not at all useful), over 2 (slightly useful), to 3 (somewhat useful) and 4 (very useful). 
Repeated measures ANOVAs did not indicate any significant difference between the lessons.

\subsubsection*{Students learned new things in the Cybercompass lessons.}
An important indicator for deciding whether to include topics of everyday cybersecurity in a compulsory ISC course, is whether it provides students with new insights.
Subsequently, we asked students for each lesson whether they had learned something new (answered on a yes/no scale plus an open-ended text field for detailing the new learning).
\textit{Passwords} was the lesson with the least amount of new learning (34\%), while \textit{Security of devices} (58\%) and \textit{Self-defense} (57\%) ranked moderately, and \textit{Secure communication} (83\%) and \textit{Incident reporting} (83\%) provided the most new insights.
Cochran's Q test revealed that the learning between the lessons differed significantly, $Q(df=4, N=130)=123.42$, $p<.001$, $\eta^2_{Q}=.24$.
In particular, Bonferroni-corrected post-hoc tests (McNemar~\cite{mcnemar1947note}) indicated that \textit{Passwords} differed from all other lessons ($p<.001$) and that \textit{Security of devices} and \textit{Self-defense} both differed from \textit{Secure communication} and \textit{Incident reporting} ($p<.001$).
Open-ended answers revealed that the most salient new things for students concerned: the 3-2-1 back-up rule, anti-theft tracking and device encryption, the breakability and creation of passphrases, the security of public Wi-Fi and the use of VPN in that context, the prevalence of phishing at our university, the possibility to use the file-sharing option in the university information system, the Eduroam configuration tool, the university VPN, the possibility of obtaining a personal certificate for e-mail encryption and signing, and the incident reporting process and contact point at our university.

\subsubsection*{Cybercompass changes the view of students on everyday cybersecurity.}
A non-negligible share of students (42\%) reported that the Cybercompass changed their view on education and preparedness in everyday security. 
When asked how the course changed their view, many indicated that it raised awareness and encouraged them to take action, as illustrated by the following quotes.

\begin{quote}
    \textit{I gained an overall view on everyday security issues, and it made me think more about security on a daily basis.}
\end{quote}

\begin{quote}
    \textit{It changed my view on passwords, I think I [will] start using password manager more and maybe also rework my passwords.}
\end{quote}

\begin{quote}
    \textit{I will definitely check the addresses of emails more often.}
\end{quote}

\subsubsection*{Cybercompass lessons encourage action.}
The course contains different kind of activities and recommendations for the use of protective tools. 
Many students indicated that they had tried at least one of these activities or tools during the course (\textit{Security of devices}: 81\%, \textit{Passwords}: 67\%, \textit{Self-defense}: 60\%, \textit{Secure communication}: 64\%).
This was especially salient in the open-ended answers for the first three lessons, as illustrated by the following quotes:

\begin{quote}
    \textit{The good point, the article encouraged me to dig around more in my smartphone security settings.} (\textit{Security of devices})
\end{quote}

\begin{quote}
    \textit{As I was reading this lesson, I sa[i]d to myself more than once, that I have to do this. So e.g. I changed my notification preview and as I am writing this an encryption of my mobile is running.} (\textit{Security of devices})
\end{quote}

\begin{quote}
    \textit{The lesson convinced me it is a good idea to set up a password manager.} (\textit{Passwords})
\end{quote}

\begin{quote}
    \textit{I also tried the challenge with recognising phishing emails and I was not that successfull, so that surprised me.} (\textit{Self-defense})
\end{quote}

\begin{quote}
   \textit{I liked especially the interactive part -- phishing quiz which I will definitely remember for a long time.} (\textit{Self-defense})
\end{quote} 

\subsubsection*{Bonus material is highly appreciated.}
Students highly appreciated the bonus material provided at the end of each lesson.
For each lesson,
a high amount of students indicated that they had at least partially read the provided material (\textit{Security of devices}: 91\%, \textit{Passwords}: 94\%, \textit{Self-defense}: 88\%, \textit{Secure communication}: 87\%).

\subsubsection*{Students are willing to recommend the Cybercompass to others.}
As IT knowledgeable users are an important information source for cybersecurity advice for people without IT background~\cite{redmiles2016learned}, we asked students whether they would recommend the course to people in their circles. 
71\% said that they would recommend it to members of their family, 68\% to their non-university friends, 54\% to fellow students, and even 28\% would recommend it to work colleagues.

\subsection{Challenges}

\subsubsection*{Lessons vary in comprehensibility and difficulty.}
We had students evaluate the comprehensibility and difficulty of each lesson. Both were rated on a four-point scale from 1 (not at all), over 2 (slightly) to 3 (somewhat) and 4 (very). 
The lessons were perceived as \textit{very comprehensible} on average, with only \textit{Secure communication} showing a tendency towards \textit{somewhat comprehensible} (\textit{Security of devices}: $M=3.65$, $SD=.58$; \textit{Passwords}: $M=3.70$, $SD=.64$; \textit{Self-defense}: $M=3.58$, $SD=.67$, \textit{Secure communication}: $M=3.48$, $SD=.71$, \textit{Incident reporting}: $M=3.74$, $SD=.68$). 
A repeated measures ANOVA showed that the comprehensibility differed significantly between the lessons, $F(3.36, 432.88)=8.03$, $p<.001$, $\eta^2_{part.}=.06$. In particular, Bonferroni-corrected post-hoc tests revealed that the \textit{Secure communication} lesson was perceived as less comprehensible than the \textit{Security of devices} ($p=.02$), the \textit{Passwords} ($p=.001$), and the \textit{Incident reporting} lesson ($p=.001$). 

On average, the lessons were perceived as \textit{not at all difficult} with only \textit{Secure communication} having a tendency towards \textit{slightly difficult} (\textit{Security of devices}: $M=1.38$, $SD=.56$; \textit{Passwords}: $M=1.22$, $SD=.41$; \textit{Self-defense}: $M=1.47$, $SD=.65$, \textit{Secure communication}: $M=1.66$, $SD=.78$, \textit{Incident reporting}: $M=1.11$, $SD=.34$).  
A repeated measures ANOVA showed that the difficulty differed between the lessons, $F(3.30, 425.00)=25.60$, $p<.001$, $\eta^2_{part.}=.17$. 
In particular, Bonferroni-corrected post-hoc tests revealed that \textit{Secure communication} was the most difficult lesson, differing significantly from \textit{Security of devices} ($p<.001$), \textit{Passwords} ($p<.001$), and \textit{Incident reporting} ($p<.001$).

\subsubsection*{Most students have not heard of the course before.} 
A huge majority (91.7\%) of participants had not heard about the Cybercompass before. 
This is surprising, given that the course is advertised through several channels within the university environment.
Those who had heard about the course indicated that this was through the university social media channels (LinkedIn and Facebook), the university information system news section, physical bulletins, the website of our school of computer science, a classmate, and an external website.

\subsubsection*{Information on password managers and 2FA is insufficient.}
As part of the \textit{Passwords} lessons, learners are presented with a tutorial on how to install a specific password manager. 
Moreover, in the bonus section of that lesson, there is a brief section on how to set up 2FA in three popular online services.
In the open-ended answers that followed the Likert-scale overall rating for each lesson, many students criticized that the provided information on those two topics is insufficient, as illustrated by the following quotes.

\begin{quote}
    \textit{I'm not rating [the lesson] `very positive` because I feel like offline password managers should be mentioned.}
\end{quote}

\begin{quote}
    \textit{I've never used a password manager and what would really help me to convince me is recommending a free (or very cheap) password manager, [...]
    explaining the process of retrieving the passwords in case I loose access to my account, explaining how and where are the passwords stored [...] 
    }
\end{quote}

\begin{quote}
    \textit{Maybe I would emphasize the use of the two factor authentication more, I think it should be a standard these days, not something 'more'.}
\end{quote}

\begin{quote}
    \textit{I think that at some class in [the ISC course], we were told that having SMS as second factor is not that secure.}
\end{quote}

\section{Discussion and Conclusion}

We evaluated an online cybersecurity awareness course with 138 computer science undergraduates -- future IT professionals. The students valued the course highly, reporting new learning, changes in their perspectives, and transfer to practice. At the same time, they suggested suitable improvements to the course. Evaluating the course yielded lessons that we processed into recommendations to help designers of similar courses and security educators. 

\subsection{Discussion}

Students learned the most in the lessons on \textit{Secure communication} and \textit{Incident reporting}.
The \textit{Secure communication} lesson familiarized students with the university IT services. 
Although students usually take the ISC course in their second year, many were not aware of the offered variety of services for secure communication.
Similarly, many students did not know where to report an incident and what the reporting process should look like.
For quite a few of the students (42\%), the Cybercompass even influenced their view on everyday cybersecurity.
This indicates that the course constitutes a valuable resource for increasing awareness. 

Students' comments included helpful ideas that we can incorporate into future versions of the course.
For instance, students wished to see more about 2FA and a broader range of password managers covered. 
Similarly, students' ratings indicated that the \textit{Secure communication} lesson is slightly less comprehensive and more difficult than the other lessons and thus needs to be simplified.
This is in line with observations from related work, which assert that especially email encryption and signing are notoriously hard to understand for users~\cite{whitten1999johnny,sheng2006johnny,ruoti2015johnny}. 

Ratings of the course were mostly positive, yet few students had heard about it before despite the university-wide dissemination efforts. 
This suggests that this kind of educational material needs different and additional promotion channels.
We believe that including the course in first-year introductory lectures would be a good way to achieve a broader reach. 
Therefore, we plan to investigate how to convince other educators to have the course in the first year. 

Our results further showed that students are willing to disseminate the course among fellow students, family members, friends, and colleagues.
Subsequently, students could act as ``cybersecurity advocates: individuals who encourage positive change by promoting and providing guidance on security best practices and technologies''~\cite{haney2017work}.
As such, they are indispensable for increasing security in different ecosystems, even outside the university. 

\subsection{Limitations}
As we evaluated the Cybercompass with computer science undergraduates, results can only be generalized to this kind of population. 
Future work can evaluate the course with different kind of populations such as students from other faculties and employees of the university.
As the data was coded by an experienced analyst, we did not calculate the inter-rater agreement. Thus, the qualitative results should be generalized with caution.

\subsection{Recommendations for Course Designers}
\label{sec:recommendations}

\paragraph{Use the Cybercompass as an inspiration.}
The positive evaluation and successes reported in Section~\ref{sec:insights-successes} indicate that the course targets the right topics.
As such, visit the course website and take it as an inspiration.

\paragraph{Encourage action.}
Activities and tools included into the Cybercompass, such as a phishing quiz, a tutorial for getting a password manager, and a hint to review smartphone settings were welcomed by many students. 
Similarly, the literature asserts that security advice should be actionable~\cite{redmiles2020comprehensive}. 
Include clear calls to action into your course if you want to make a difference to people's security habits.

\paragraph{Include bonus material for curious users.}
Our student sample appreciated the bonus course material.
If your audience is diverse as ours (coming from various schools and institutes, plus staff and public), add extra material for curious users. 

\paragraph{Evaluate dissemination channels and measure reach.}
Although the Cybercompass was widely promoted on university online channels, its reach seems to be limited as less than 10\% of our sample had heard of the course before. 
If you are designing a similar course, make sure to reach out to the intended audience in more creative ways than we did. 
Additionally, try to measure the return rate on different channels to evaluate the effectiveness of each channel.

\paragraph{Update the course regularly.} 
Everyday cybersecurity is prone to change. 
With evolving platforms and tools, recommendations should be adjusted at least once per year.
As in our case, we appreciate that the computer science students hinted us towards including more information on 2FA and password managers. 
We will stay in touch with experts and users to identify trending topics. 

\subsection{Recommendations for Security Educators}
\label{sec:recommendations2}

\paragraph{Think of your computer science students as future cybersecurity advocates.}
Our results show that an overwhelming majority of computer science students is willing to recommend the Cybercompass to others.
Providing students with a unified source of everyday security advice does not only serve them, but is also likely to increase information security awareness at the university and in society.
As such, think of your students as future advocates -- even the advanced who do not learn something new in that course can pass on the knowledge to others.

\paragraph{Consider including topics of everyday cybersecurity into your information security courses.}
Our results show that including the Cybercompass into a compulsory introductory ISC course yielded positive experiences among students.
Students learned new things in all lessons, reported increased awareness, and were encouraged to take action.
Even if curricula constraints are tight, consider including material of everyday cybersecurity -- at least the \textit{Secure communication} and \textit{Incident reporting} lessons, which showed to yield the most new learning.

\begin{acks}
This research was supported by \grantsponsor{ERDF}{ERDF}{} project \textit{CyberSecurity, CyberCrime and Critical Information Infrastructures Center of Excellence} (No. \grantnum{ERDF}{CZ.02.1.01/0.0/0.0/16\_019/0000822}). We would like to thank Martin Ukrop for help with the questionnaire deployment, and the students who participated in the survey. Furthermore, we would like to thank Adam Skrášek and Elizabeth Stobert for helpful comments during the pilot testing.
\end{acks}


\bibliographystyle{ACM-Reference-Format}
\bibliography{bibliography}

\end{document}